\documentclass[twocolumn,showpacs,preprintnumbers,amsmath,amssymb]{revtex4}

\usepackage{graphicx}
\usepackage{dcolumn}
\usepackage{bm}

\usepackage{graphicx}
\usepackage{epsfig}

\begin{document}

\title{On exact solutions to the scalar field equations in standard cosmology}  

\author{Marco A. Reyes}
\email{marco@fisica.ugto.mx}

\affiliation{Instituto de F\'{\i}sica, Universidad de Guanajuato,
Loma del Bosque 103, Frac. Lomas del Campestre, C.P. 37150 Le\'{o}n,
Guanajuato, M\'{e}xico.
}

\begin{abstract}
A simple algebraic method to obtain exact solutions to the scalar field
equations in spatially flat FRW cosmology is derived.  The field potential
fuction is reduced to two terms which can be used to determine some 
characteristic inflationary variables. To demonstrate the method, exact
solutions resembling typical potentials of general interest are worked out. 
Finally, it is shown that this method of solution is also applicable to a
perfect fluid in the same metric, for a curvature $k\geq 0$ and fixed $\gamma$.
\end{abstract}

\maketitle

\section{Introduction}
It is a common issue in Cosmology  to make use of scalar fields $\phi$ as the
responsible agents of some of the most intriguing aspects of our universe (see
\cite{inflation} and references therein). Just to mention a few, we find that
scalar fields are used as the inflaton which seeds the primordial perturbations
for structure formation during an early inflationary epoch; as the cold dark
matter candidate responsible for the formation of the actual cosmological
structure and as the dark energy component which seems to be driving the
current accelerated expansion of the universe.  The key feature for such
flexibility  of scalar fields (spin-$0$ bosons) is the freedom one has to
propose a \textit{scalar potential} $V(\phi)$, which encodes in itself the (non
gravitational) self-interactions among the scalar particles. Recently, scalar
fields coupled to gravity (in a FRW background) have also appeared in
connection to the so called string theory landscape \cite{landscape}, where the
scalar potential $V(\phi)$ is usually thought of as having many valleys, which
represent the different vacua solutions. The hope is that the statistics of
these vacua could explain, for example, the smallness of the cosmological
constant (the simplest candidate for dark energy).

This coupling of gravity with scalar fields is what is known as scalar field
cosmology. In order to construct this model we start with the line element for a
homogeneous, flat and isotropic universe, the  Friedmann-Robertson-Walker (FRW)
metric 
\begin{equation}
ds^2= -dt^2 + e^{2\alpha(t)}\left[dr^2
+r^2 d\Omega^2 \right] , 
\label{frw}
\end{equation}
where $ a(t)=e^{\alpha(t)}$ is the scale factor. The classical evolution is
obtained by solving Einstein's field equations. Using the energy momentum tensor
of a scalar field we get the Einsten-Klein-Gordon equations
\begin{eqnarray}
H^2 = \rm \frac{8\pi G}{3} \left (\frac{1}{2}\dot{\phi}^2+V(\phi)\right ), 
\nonumber \\
\ddot{\phi}+3H\dot{\phi} = -\frac{\partial V(\phi)}{\partial\phi} \ ,
\hspace*{5mm}
\label{fried}
\end{eqnarray}
where $ H=\dot{a}/a$.
Also, $H$ satisfies that $\dot H$=$-\dot\phi^2${\small /2}.
As stated at the beginning, by 
choosing appropriate  {potentials}, different aspects of the universe can be
modeled.   In the following we shall use $8\pi G$=1.

Because of the non linear character of Einstein's equations,  exact solutions to
the scalar cosmology  equations can only be found using simple methods for very
particular potentials (constant and exponential). But for more
appealing potentials, that explain some of the properties of our universe, exact
solutions can not be found. In order to find exact solutions for complicated
potentials one must use elaborate techniques, like Chimento and Jakubi's
solution for a perfect fluid \cite{chimen}. 
However, the need for an exact solution to the problem has been surpassed by the
fact that what matters most in the model is the dynamical aspects of the
field, and therefore, the approximations made using the {\it slow roll}
method have prevailed over the need to find such exact solutions. 
This approximation essentially means that the dynamical terms of the scalar
field, denoted by  $\frac{1}{2}\dot\phi^2$ and by $\ddot\phi$ in
eqs.(\ref{fried}), are neglected with respect to the other terms in the
equations. This permits us to write down the physical quantities in terms of the
slow roll parameters $\epsilon$ and $\eta$. Also the use of numerical in
connection with dynamical systems is common to complete the
solution\cite{luislidd}.

Despite the success of the {\it slow roll} approximation, it is clear that
having a more general method to find exact solutions to eqs.(\ref{fried}) may
allow us to make a better description of the  phenomenology of this particular
model, and would give a direct way to probe the exactness of the numerical
algorithms extensively used.

The present work is in this direction, to develop a simple algebraic method to
find exact solutions for a wide variety of scalar field potentials. The paper
is arranged as follows: in Section II the method is presented for an arbitrary
scalar field potential, followed by a disscusion on dynamics in  Section III. 
A few examples of general interest are presented in Section IV, and the
connection to the exact solutions of scalar field equations in perfect fluid is
made in Section V.  Conclusions and outlook are given in Section VI. 


\section{The algebraic method}

The system of coupled eqs.(\ref{fried}) has three unknowns,
$\phi(t)$, $V(\phi)$ and $a(t)$.  To solve it, one of these variables has to be
given a  priori.  It is customary to look for the solution for a given
$V(\phi)$, but as it is known, very few potentials are known to lead to exact
solutions.

In this work, it is proposed to reduce the equations to a simpler form first,
which helps to solve them exactly.  In order to do this, let us consider the
function $V_a(\phi)$ defined as 
\begin{equation}\label{va-vf} 
V_a(\phi) \equiv V(\phi)+\frac{1}{2}\dot\phi^2
\end{equation}
Now, with the change of variable $dt\!=\!d\phi/\dot\phi$, one obtains 
\begin{equation}\label{v2} 
\frac{d V_a}{d\phi}=\frac{d V}{d\phi} + \ddot\phi
\end{equation}
Hence, eqs.(\ref{fried}) can now be rewritten as
\begin{eqnarray}
&& 3\left( \frac{\dot a}{a} \right)^2 = V_a
\label{frw2a}
\\
&& 3\left( \frac{\dot a}{a} \right) \dot\phi = -\frac{dV_a}{d\phi}
\label{frw2b}
\end{eqnarray}

These equations are now easily solved, and the function $V_a(\phi)$ ($>0$) shows
up as the main part of the potential function, driving the dynamics of the scale
factor.  To solve them, note that eq.(\ref{frw2a}) defines $\dot a/a$ as a
function of $\phi$, $H(\phi)$, which when inserted into eq.(\ref{frw2b}), 
gives the scalar field $\phi(t)$ as a function of $t$, at least in quadratures
\begin{equation}\label{cuadr}
- 3 H(\phi) \left( \frac{d V_a}{d\phi} \right)^{-1} d\phi =  dt
\end{equation}

Finally, inserting $\phi(t)$ into eqs.(\ref{va-vf}) and (\ref{frw2a}) gives
$V(\phi)$ and $a(t)$, respectively, and the solution is completed.  

Obviously,
one could simply have begun by giving $H$=$H(\phi)$, but it is usually desired to
have some description of the potential instead, and for this reason it is
preferable to give $V_a(\phi)$.
One could also use $H(t)$ to determine $\phi(t)$, since 
\begin{equation}
\frac{1}{2}\dot\phi^2 = -\frac{d H(t)}{dt}
\end{equation}
implies that
\begin{equation}
\Delta\phi(t) = \pm\int\sqrt{-2 \frac{d H(t)}{dt}} ~ {dt}
\end{equation}
and since $V_a\!=\!3H^2(t)$, a complete knowledge of $H(t)$ fully determines
the solution to the problem.  For example, if $H(t)=\alpha/t$, then 
$V(\phi)\!=\! (3\alpha^2-\alpha) \exp[-2\Delta\phi/\sqrt{2\alpha}]$ and 
$\Delta\phi(t)=\sqrt{2\alpha}\ln t$. 
The difference when we proceed this way is that we lose control of the form of
the potential $V(\phi)$, and the solutions, for example, may not comply with
slow roll.\cite{ellis}


\section{Dynamical variables}

The potential function $V(\phi)$ in eq.(\ref{va-vf}) consists now of two
positively defined competing terms, $V_a$ and $\dot\phi^2/2$ (see the solution
in ref.\cite{feb08}), the second term being neglected in the {\it slow roll}
approximation.  In the limit $\dot\phi \to 0$, we find that  
\begin{equation}\label{vlim}
\lim_{\dot\phi \to 0} V(\phi) = V_a(\phi)
\end{equation}
and the scalar field potential is just $V_a$.

Finding two competing terms to drive the dynamics of the system has a quantum
mechanics equivalent.  For the phase space dynamics of the Schr\"odinger
equation in one dimension,  in terms of the wave function $\psi(x)$ and its
derivative with respect to $x$, $\psi'(x)$ \cite{tesismaes}, the potential
function $V(x)$ appears as the driving term of the dynamics, 
competing with the decelerating term defined by the energy E of the system,
and leading to the bifurcation trajectories in phase space \cite{tesismaes}.  
Moreover, the similarities of the scalar field equations with a non-linear
Schr\"odinger problem can be further realized \cite{schrod}.

One can use $V_a$ to determine some characteristics of the dynamics of the
system.  For example, from eq.(\ref{frw2a}) we have that
\[
\frac{\dot a}{a} = \pm \sqrt{\frac{V_a}{3}}
\]
and from $\dot H$=$\ddot a/a-(\dot a/a)^2$=$-\dot\phi^2/2$, that
\[
\frac{\ddot a}{a} = \frac{1}{3}V_a - \frac{1}{3}\dot\phi^2
\]
Hence $\ddot a$ has the same sign as $a$ when $\dot\phi\to 0$.

To end this section, let us determine the parameters $\epsilon$ and $\eta$ 
used in the literature to quantify the inflation regime.  Here, we can define
them as functions of the potential terms 
\begin{equation}\label{epsilon}
\epsilon(t) \equiv \frac{d~}{dt} \frac{1}{H} = 3\frac{T_\phi}{V_a}
\end{equation}
\begin{equation}\label{delta}
\eta(t) \equiv -\frac{1}{2H} \frac{\ddot H}{\dot H} = 
-3\frac{\dot T_\phi}{\dot V_a} = 
-3\frac{T_\phi^\prime}{V_a^\prime}
\end{equation}
where, to simplify, we wrote $T_\phi$ for $\frac{1}{2}\dot\phi^2$, and the primes
denote differentiation with respect to $\phi$.
Hence, $\epsilon$ and $\eta$ are the relative strenghts of the two potential
terms, and of their derivatives.
In the literature, a value of $\epsilon\!=\!2$ defines the radiation era, while
$\epsilon<1$ defines an inflationary epoch \cite{libro}.  Therefore, $V_a$ is
the dominant term of the potential function during inflation, a condition that
clarifies its appearance in the solution.


\section{Few examples of general interest}

Now, let us consider some examples of general interest.  Keeping in mind that
$V_a$ is the main part of the solution, the examples are labeled accordingly.

\subsection{\mbox{\boldmath $V_a=\lambda \phi^2$}}

The $\lambda \phi^2$ potential is one of the basic functions used in the
literature \cite{lamf2}, and it is a basic potential in standard cosmology
\cite{luismayra}. 

In this case, $H^2$=$\lambda\phi^2/3$, and $dV_a/d\phi$=$2\lambda\phi$. 
Therefore, from eq.(\ref{cuadr}) it is easy to find that
\begin{equation}\label{vlf2-1}
\Delta \phi(t) = \pm 2 \sqrt{\frac{\lambda}{3}} ~ \Delta t
\end{equation}
Hence, $\dot\phi$ is constant.  Letting $\phi_0=\phi(t_0=0)=0$, and using
eqs.(\ref{va-vf}) and (\ref{frw2a}) we get 
\begin{equation}\label{vlf2-3}
V(\phi) = \lambda\phi^2 - 2 \frac{\lambda}{3}
\end{equation}
\begin{equation}\label{vlf2-2}
a(t) = a_0 e^{-\frac{\lambda}{3} t^2}
\end{equation}
Obviously, one would be tempted to pick $\lambda<0$ in order to make $a(t)$ a
growing function of $t$, but that would make $\phi(t)$ an imaginary
function of $t$.


\subsection{\mbox{\boldmath $V_a=\lambda \phi^4$}}

The $\lambda \phi^4$ potential is also one of the basic functions used in the
literature \cite{lamphi4a}, and in cosmological models \cite{lamphi4b}.  
The solution is
\begin{equation}\label{vlf4-1}
\phi(t) = \phi_0 e^{\pm 4 \sqrt{\frac{\lambda}{3}} t}
\end{equation}
\begin{equation}\label{vlf4-3}
V(\phi) = \lambda \phi^4 - \frac{8}{3}\lambda \phi^2
\end{equation}
\begin{equation}\label{vlf4-2}
a(t) = a_0 \exp\left\{ -\frac{\phi_0^2}{8}
 e^{\pm 8 \sqrt{\frac{\lambda}{3}} (t-t_0)} \right\}
\end{equation}
Notice the appearance of a double exponential in $a(t)$.  In this case, the
radiation era happens when $\phi\!=\!2$, and inflation comes when
$\phi\!>\!2\sqrt{2}$.


\subsection{\mbox{\boldmath $V_a=\lambda \phi^{2n}$, $n>2$}}

The cases with integer $n>2$, all have the same kind of
solution.  Although this is not a very common potential in the literature
\cite{lamphi2n}, it is solved here to compare with the two previous cases.

\begin{equation}\label{vlf2n-1}
\phi(t) = 
\left[ \phi_0^{2-n} \pm 
2n (n-2) \sqrt{\frac{\lambda}{3}} ~ (t-t_0) \right]^{\frac{-1}{n-2}}
\end{equation}
\begin{equation}\label{vlf2n-3}
V(\phi) = \lambda \phi^{2n} - \lambda \frac{2 n^2}{3}\phi^{2(n-1)}
\end{equation}
\begin{equation}\label{vlf2n-2}
a(t) = a_0 \exp\left\{ \frac{-1}{4n} \left[ \phi_0^{2-n} \pm
2n (n-2) \sqrt{\frac{\lambda}{3}} (t-t_0) \right]^{\frac{-2}{n-2}} \right\}
\end{equation}
%


\subsection{\mbox{\boldmath $V_a=V_o e^{-\alpha \phi}$}}

One of the few problems exactly solved in the literature, is that of an
exponential potential function \cite{expo}.  
In this case, the solution found with the method described above gives,
for $\phi_0=0$,
\begin{equation}\label{vexp-1}
\phi(t) = \frac{2}{\alpha} \ln \left( 1+\frac{\alpha^2}{2}
\sqrt{\frac{V_o}{3}} ~ t  \right)
\end{equation}
\begin{equation}\label{vexp-3}
V(\phi) = V_o  \left(1-\frac{\alpha^2}{6}\right) e^{-\alpha\phi}
\end{equation}
\begin{equation}\label{vexp-2}
a(t) = a_0 \left( 1+\frac{\alpha^2}{2}\sqrt{\frac{V_o}{3}} 
t \right)^{\frac{2}{\alpha^2}} 
\end{equation}
The reader can recognize now that this is the solution used in the literature,
where the two terms of $V(\phi)$ are grouped together, leaving the
essence of the competing term $V_a$ out of the picture.


\subsection{\mbox{\boldmath $V_a=V_o \left( \cosh(\beta \phi)-1 \right)$}}

Another potential used in the literature is that of a hyperbolic cosine
\cite{matos}.  The solution found here, with $\phi_0$=0, is 
\begin{equation}\label{cosh-1}
\phi(t) = \frac{-2}{\beta} \mbox{arcsinh} \left( \tan \left( 
\sqrt{\frac{V_o}{6}} \beta^2 t \right) \right)
\end{equation}
\begin{equation}\label{cosh-3}
V(\phi) = V_o \left( \cosh(\beta \phi)-1 \right) -
V_o \frac{\beta^2}{6} \left( \cosh(\beta\phi) +1 \right)
\end{equation}
\begin{equation}\label{cosh-2}
a(t) = a_0 \cos^{\frac{2}{\beta^2}}\left( \sqrt{\frac{V_o}{6}}\beta^2 t  
\right) 
\end{equation}
In this case, $\epsilon(t) = (\beta^2/2) \coth^2(\beta\phi/2)$.


\subsection{A Morse type potential}

The previous examples were related to potentials used in the literature.  They
may, however, not be able to comply with the desired characteristics of the
model.  Nevertheless, one can now generate a variety of solutions to
eqs.(\ref{fried}) that may be more useful for that purpose. 

If, for example, one would like to find a potential function as smooth as the
Morse potential $V(x)=A(e^{2\alpha x}-2e^{\alpha x})$ with the method described
here, one can begin by letting 
\[
V_a = \alpha^2 \left( e^{\beta \phi} -1 \right) ^2
\]
Then, we will have that
\[
\phi(t) = 
-\frac{1}{\beta} \ln \left( 1\pm \frac{2\beta^2\alpha}{\sqrt{3}}t \right)
\]
and
\[
V(\phi) = \alpha^2 \left[ \left( 1-\frac{2\beta^2}{3} \right) e^{2\beta\phi}
-2e^{\beta\phi} +1 \right]
\]
By requiring, for example, that $\beta^2=\frac{3}{4}$, we end up with the
potential  
\begin{equation}\label{morse}
V(\phi) = \alpha^2 \left[ \frac{1}{2} e^{\sqrt{3}\,\phi}
-2e^{\sqrt{3}\,\phi/2} +1 \right]
\end{equation}
This potential has its minimum at $\phi_m=\frac{2}{\sqrt{3}}\ln 2$ with a value
of $V(\phi_m)=-\alpha^2$, and with $V''(\phi_m)>0$.  It is sketched in Fig.(1). 
The value for $\phi$$\to$$-\infty$ is $\alpha^2$.  Thus, the smaller the $\alpha$ the
smoother the curve. 
\begin{figure}[h!]
\epsfig{figure=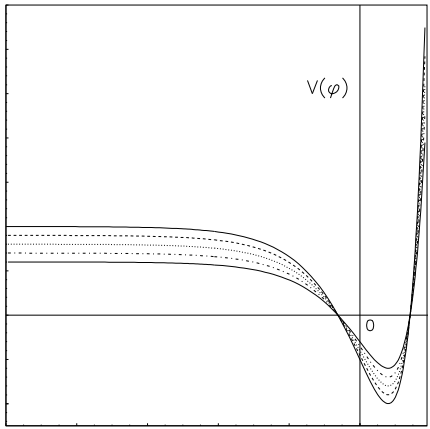, height=7cm}
\vspace*{-3mm}
\caption{A Morse type potential, the kind that may satisfy the {\it slow roll}
conditions.}
\label{fig1}
\end{figure}

This kind of potential may be more capable of satisfying the {\it slow roll}
conditions.  Moreover, by requiring the lower sign in the previous equations, we
find that $a(t)$ is
\begin{equation}
a(t) = a_o \, e^{\alpha t/\sqrt{3}} \left( 1-\frac{2\beta^2\alpha}{\sqrt{3}}
\right)^ {\sqrt{3}/2\beta^2\alpha} 
\end{equation}
therefore allowing inflation.


\section{Solutions for a perfect fluid with $k\geq 0$ and fixed $\gamma$}

The case considered here of a scalar field in a spatially flat FRW metric is
the simplest one in the model.  If the equations were linear, one
would expect to infer from this method solutions to more
complex systems.  Since this is not the case, that is very difficult
to achieve.  However, one can still find some cases where the method
of Section II may be useful.

Let us begin by noting that since
\begin{eqnarray}
\frac{d~}{dt} \! \! \left( 3H^2-\frac{1}{2}\dot\phi^2-V(\phi) 
+ \mbox{\it const.} \right) =  &&  \\
- \dot\phi ~ \left( 3H\dot\phi +\ddot\phi + \frac{dV(\phi)}{d\phi} \right) &&
\end{eqnarray}
then any pair of equations of the form 
\begin{eqnarray}
\ddot\phi + 3H\dot\phi +\frac{dV(\phi)}{d\phi} = 0  \nonumber \\
3H^2 = \frac{1}{2}\dot\phi^2 + V(\phi) + F(t) + \mbox{\it const.}  \nonumber 
\end{eqnarray}
will have the same solutions as the spatially flat case, provided that
$F(t)\equiv 0$. We shall denote the integration constant appearing in these
equations as $\Lambda$ \footnote{Here we only concentrate on the algebraic
solutions, without further implications on the phenomenology of the model.}.

If we had included $\Lambda$ in eqs.(\ref{fried}), we
would have found  that in order to have a potential
$V(\phi)$=$\lambda\phi^2$, $\Lambda$=$-2\lambda/3$, and that a value of
$\Lambda=16\lambda/9$ would have been found for the potential
$V(\phi)=\lambda(\phi^2-4/3)^2$. 

Now, the scalar field equations for a perfect fluid in the
FRW metric, as used by Chimento and Jakubi \cite{chimen}, have the structure
given above.  Those equations are 
\begin{eqnarray}
&& \ddot\phi + 3H\dot\phi +\frac{dV(\phi)}{d\phi} = 0  \label{pf1} \\
3H^2 &=& \frac{1}{2}\dot\phi^2 + V(\phi) + \rho - 3\frac{k}{a^2} +\Lambda 
\label{pf2} 
\end{eqnarray}
where the energy density and pressure of the fluid are given by 
$\rho=\rho_0/a^{3\gamma}$ and $p=(\gamma-1)\rho$, respectively, with 
$0\leq\gamma\leq 2$.

Hence, we have that any solution to eqs.(\ref{fried}) is 
a solution for the pair of coupled eqs.(\ref{pf1}-\ref{pf2}), provided that
\begin{eqnarray}
&& k\geq 0 \ ,\\
&& \rho(t)=3\frac{k}{a^2} \ , \\
&& \gamma = \frac{2}{3}
\end{eqnarray}

Notice that these conditions also satisfy that
\begin{equation}
\dot H = -\frac{1}{2}\dot\phi^2-\frac{1}{2}\gamma\rho+\frac{k}{a^2}
= -\frac{1}{2}\dot\phi^2
\end{equation}
and the connection between solutions is complete.

In the literature, a value of $\gamma=2/3$ in cosmology is related to a static
universe.  Here, this value appears even in solutions where $H(t)$ is not
constant. 

\vspace*{3mm}

\section{Conclusions and Outlook}

In this work, the solution to eqs.(\ref{fried}) proceeded by giving
an ansatz for part of the potential function, which was then related to the
parameters that define the inflation regime.  One can also start by proposing 
the field $\phi(t)$ and from there obtain the whole solution.

If we proposed $\phi$ as a polynomial function of $t$, $\phi\!=\!\sum a_n
t^{n}$,\cite{lidd-n} 
then we would find that the highest exponent would give $\dot\phi\propto
t^{n-1}$, then $H\propto t^{2n-1}$, and $V(t)\propto t^{4n-2}$.  If we worked out
the case $n=1$, we would arrive to the solution given in Section IV.A, with the
leading term of $V(\phi)$ proportional to $\phi^2$. However, for $n\gg 1$, we
would find that $V\sim \phi^4$, therefore leading to the roots of the
$\lambda\phi^2$ and $\lambda\phi^4$ models.

In conclusion, we have given here an algebraic method to find a large variety
of solutions to the scalar field equations in spatially flat FRW metric, which 
may be useful in more complex systems.  Hopefully, it will
give a deeper insight into the phenomenological aspects of the scalar field
model.  


\section{Acknowledgments}

I would like to thank M. Sabido, W. Guzm\'an, J. Socorro and L. Ure\~na
for very elucidating conversations.

\end{document}